\documentclass[a4paper,fleqn,usenatbib]{mnras}
\usepackage[dvips]{graphicx}
\usepackage{amssymb}
\usepackage{amsmath}
\usepackage{graphicx,color}
\usepackage{amsfonts}
\usepackage{bm}
\usepackage{cancel}
\usepackage{comment}

\newcommand\be{\begin{equation}}
\newcommand\ee{\end{equation}}

\newcommand\e{\mathrm{e}}

\allowdisplaybreaks[4]

\title{Chandrasekhar Mass Limit of White Dwarfs in Modified Gravity}
\author[Astashenok, Odintsov \& Oikonomou]{Artyom V. Astashenok$^{1}$, Sergey D. Odintsov$^{2,3}$, Vasilis K. Oikonomou$^{4}$\\
\small $^{1}$Institute of Physics, Mathematics and IT, I.
Kant Baltic Federal University, 236041 Kaliningrad, Russia\\
\small $^{2}$ICREA, Passeig Luis Companys, 23, 08010 Barcelona, Spain \\
\small $^{3}$Institute of Space Sciences (ICE,CSIC) C. Can Magrans s/n,
08193 Barcelona, Spain\\
\small $^{4}$Department of Physics, Aristotle University of
Thessaloniki, 54124, Thessaloniki, Greece\\}

\begin{document}

\label{firstpage}
\pagerange{\pageref{firstpage}--\pageref{lastpage}} \maketitle

\maketitle

\begin{abstract}
We investigate the Chandrasekhar mass limit for white dwarfs in
various models of $f(R)$ gravity. Two equations of state for
stellar matter are used: simple relativistic polytropic equation
with polytropic index $n=3$ and the realistic Chandrasekhar
equation of state. For calculations it is convenient to use the
equivalent scalar-tensor theory in the Einstein frame and then to
return in the Jordan frame picture. For white dwarfs we can
neglect terms containing relativistic effects from General
Relativity and we consider the reduced system of equations. Its
solution for any model of $f(R)=R+\beta R^{m}$ ($m\geq 2$,
$\beta>0$) gravity leads to the conclusion that the stellar mass
decreases in comparison with standard General Relativity. For
realistic equations of state we find that there is a value of the
central density for which the mass of white dwarf peaks.
Therefore, in frames of modified gravity there is lower limit on
the radius of stable white dwarfs and this minimal radius is
greater than in General Relativity.

\end{abstract}


\section{Introduction}

Modified gravity in its various forms
(\citealp{reviews1,reviews2,reviews4,book,reviews5,reviews6,dimo})
describes successfully and in a minimal way the late-time
acceleration of the universe (\citealp{Perlmutter, Riess1, Riess2}) and
in many cases, the unification of the inflationary era with the
early-time acceleration is possible (\citealp{Nojiri:2003ft}). Although
the most successful model for the cosmological acceleration is the
$\Lambda$-Cold-Dark-Matter ($\Lambda$CDM) model, but it has
several shortcomings from the theoretical physics viewpoint.
Primarily one needs to explain the so called cosmological constant
problem i.e. the very large discrepancy between observed value of
$\Lambda$ term and its value predicted by any quantum field
theories (\citealp{Weinberg}). Another way to describe cosmological
acceleration in frames of General Relativity (GR) is the
introduction of a scalar field. Analysis of the Planck
observational data leads to conclusion that such field may be
phantom field with negative kinetic terms, since the dark energy
equation of state (EoS) parameter is allowed to have values
marginally smaller than $-1$. Phantom fields are very problematic
from a quantum field theory perspective.

In modified gravity we can explain not only data based on standard
candles, but also microwave background anisotropy (\citealp{Spergel}),
shear due to the gravitational weak lensing (\citealp{Schmidt}), data
about absorption lines in Lyman-a-forest (\citealp{McDonald}) and other
without cosmological constant or phantom scalars.

However, when considering modifications
of GR, a holistic approach compels to consider not only possible
manifestations of such theories at a cosmological level, but also
at an relativistic astrophysical level, also because strong
gravitational regimes could be considered if GR is the weak field
limit of some more complicated effective gravitational theory.

In this paper we consider possible manifestations of $f(R)$
gravity in white dwarfs. Models of white dwarfs with polytropic
EoS in Palatini $f(R)$ gravity (without no extra degree of freedom
for the gravitational sector) are considered in
\citealp{Wojnar1,Wojnar2}. Early in many papers another class of
compact objects, neutron stars (NS) have been considered in
connection with modified gravity (see for example,
\citealp{Astashenok:2020qds,Astashenok:2021peo,Capozziello:2015yza,Astashenok:2014nua,Astashenok:2020cfv,Arapoglu:2010rz,Panotopoulos:2021sbf,Lobato:2020fxt,Oikonomou:2021iid,Odintsov:2021qbq,Katsuragawa:2022},
and for a recent review see \citealp{Wojnar3}). The general feature
of the solution of the modified Tolman-Oppenheimer-Volkoff
equations is that scalar curvature $R$ outside the NS doesn't drop
to zero as for for GR, but asymptotically approaches zero at the
spatial infinity (and from a calculational point of view at the
numerical infinity). The gravitational mass inside the  star's
surface, decreases in comparison with GR for same density of
nuclear matter at the center of star, but contributions to the
gravitational mass are obtained from regions beyond the surface of
the star. For a simple $R^2$ gravity effective gravitational mass
of NSs increases. This result can help to explain NS with large
mass
(\citealp{Pani:2014jra,Doneva:2013qva,Horbatsch:2015bua,Silva:2014fca,Chew:2019lsa,Blazquez-Salcedo:2020ibb,Motahar:2017blm,Oikonomou:2021iid,Odintsov:2021qbq}).
In light of the relatively recent GW190814 event, modified gravity
serves as a cutting edge probable description of nature in limits
where GR needs to be supplemented by a Occam's razor compatible
theory.

The density and the scalar curvature in central areas of white
dwarfs of course are not so large as inside NSs. But radii of
white dwarfs are two or three orders of magnitude larger and
therefore some measurable effect may appear. In Newtonian gravity
and for polytropic equation of state equations describing stars
equilibrium, give well-known Lane-Emden equation. From
calculations it follows that we can neglect relativistic effects
on Newtonian background, but is this true for possible influence
of modified gravity? The second important question is existence of
stable stars in modified gravity for realistic EoSs, for the
branch of stable stellar configurations $dM/d\rho_c>0$ where
$\rho_c$ is central density. In the case of Chandrasekhar EoS mass
increases with central density. Although for very large density
($\rho_c\sim 10^{10}$ g/cm$^3$) this EoS is not applicable it is
interesting to investigate the question about stability in
modified gravity. Since the scalar curvature is relatively small
one can expect that the function $f(R)$ can be represented as a
power series in $R$. Therefore, in first turn we should consider a
simple model of power-law gravity with an additional term $\sim
R^{l}$ to scalar curvature.

The structure of this paper is as follows: In sections II and III
we briefly consider Tolman-Oppenheimer-Volkoff equations in GR and
$f(R)$ gravity. We can neglect relativistic terms for white dwarfs
in the first case and obtain well-known Lane-Emden equation for
polytropic EoSs. For $f(R)$ gravity it is convenient to use
Einstein frame and the corresponding scalar-tensor theory.
Neglecting same relativistic terms we obtain a reduced system of
equations which is easier to numerical analysis. Then we compare
the two approaches to solve this system for a simple $R^2$ gravity
using relativistic polytropic EoS with polytropic index $n=3$.
Firstly, we can use an approximation for the scalar field. In this
case, the scalar field decreases as the density of the star
decreases and drops to zero on star surface. Stellar mass
decreases in comparison with GR. These results do not change
qualitatively if we solve the reduced system without any
approximation. Realistic Chandrasekhar EoS is considered in
Section V for $R^2$ gravity. Finally, we investigate the mass
limit for white dwarfs in another model of $f(R)$ gravity for a
polytropic EoS. Assuming a perturbative solution for the scalar
field, one can obtain the analog of Lane-Emden equation and
formulate requirements to gravity model at which the Chandrasekhar
mass limit increases or decreases.

\section{Tolman-Oppenheimer-Volkoff Equations in GR}

For relativistic non-rotating stars in equilibrium the following equations should be satisfied:
\begin{equation}\label{OV-1}
    \frac{dm}{dr} = 4\pi \rho r^2,
\end{equation}
\begin{equation}\label{OV-2}
    \frac{dp}{dr} = - (\rho + p) \frac{m+4\pi p r^3}{r^2\left(1-\frac{2m}{r}\right)}.
\end{equation}
where $\rho$ and $p$ are the density and the pressure of stellar
matter respectively. Function $m$ is the gravitational mass
enclosed in a sphere with radius $r$.  Here we use natural system
of units in which velocity of light and gravitational constant are
$c=G=1$.

Dense matter in white dwarfs can be described by simple polytropic
equation of state namely,
\begin{equation}
    p=K\rho^{1+1/n}\, ,
\end{equation}
where $K$, $n$ are constants. One should note that the pressure
and the density have the same dimensions in natural system of
units. For polytropic EoSs one can obtain simple equations for
dimensionless quantities and investigate the properties of the
solutions of the TOV equations.

Let's define the following dimensionless
functions $\theta$ and $\mu$ and the coordinate variable $x$:
$$
\rho = \rho_{c} \theta^{n},\quad m = \mu\rho_c a^3, \quad r=a x,
$$
where the length parameter $a$ is,
$$
a = \left(\frac{(n+1)K\rho_{c}^{1/n-1}}{4\pi}\right)^{1/2}.
$$
In terms of the dimensionless variables, the first equation is
equivalent to,
\begin{equation}\label{OV-4}
    \frac{d\mu}{dx} = 4\pi x^2 \theta^{n},
\end{equation}
and the second equation can be reduced to,
\begin{equation}\label{OV-3}
    \frac{1}{1+4\pi\beta\theta/(n+1)}\frac{d\theta}{dx} = -\frac{1}{4\pi} \frac{\mu+16\pi^2 (n+1)^{-1} x^{3}\beta\theta^{n+1}}{x(x-2\beta \mu)}.
\end{equation}
The dimensionless parameter $\beta$ is,
$$
\beta = \rho_c a^2 = \frac{n+1}{4\pi}K\rho_{c}^{1/n}<<1\, ,
$$
and is very small for the corresponding densities in white dwarfs.
If we consider relativistic electrons ($\rho_{c}>>10^6$ g/cm$^3$)
then $n=3$ and
$$
K=1.2435\times 10^{15}/\mu_{e}^{4/3}
$$
in CGS-system. Here $\mu_e$ is average molecular weight per one
electron. For $\mu_{e}=2$ we have that,
$$
\beta = 3.76\times 10^{-5}\left(\frac{\rho_c[\mbox{g/cm}^{3}]}{10^{7}}\right)^{1/3}
$$
If we neglect terms containing $\beta$ in Eqs. (\ref{OV-4}), (\ref{OV-3}) we obtain well-known Lane-Emden equation:
\begin{equation}
    \frac{d}{dx}\left(x^2\frac{d\theta}{dx}\right)=-x^2 \theta^{n}.
\end{equation}
For the case $n=3$, the mass of the star does not depend on the
central density and $\mu(x_{f})=25.362$ for $x_f=6.896$. This
corresponds to the Chandrasekhar limit of white dwarf mass
$M=1.456 M_{\odot}$. By taking into account relativistic terms,
the results changes negligibly.

\section{Spherically symmetric stellar configurations in f(R)-gravity}

If we consider $f(R)$ gravity, one needs to replace the standard
Einstein-Hilbert action which contains the scalar curvature $R$ by
the some function of curvature $f(R)$:
\begin{equation}\label{action}
S=\frac{1}{16\pi}\int d^4x \sqrt{-g}f(R) + S_{{\rm matter}}.
\end{equation}
Here $g$ is determinant of the metric $g_{\mu\nu}$ and $S_{\rm
matter}$ is the action of the standard perfect fluid matter.

The spherically symmetric metric for static star is
\begin{equation}\label{metric}
    ds^2= -e^{2\psi}dt^2 +e^{2\lambda}dr^2 +r^2 d\Omega^2,
\end{equation}
where $\psi$ and $\lambda$ are two independent functions of the
radial coordinate $r$.

For our purposes it is useful to consider scalar-tensor theory of
gravity which is equivalent to $f(R)$ gravity. The equivalent
action for gravitational field in the Einstein frame is
 \be S_{g}=\frac{1}{16\pi}\int
d^{4} x \sqrt{-g}\left(\Phi R - U(\Phi)\right). \ee where the
scalar field is $\Phi=f'(R)$ and the potential is
$U(\Phi)=Rf'(R)-f(R)$. Making transformation of metric
$\tilde{g}_{\mu\nu}=\Phi g_{\mu\nu}$ one write the action in the
Einstein frame, \be S_{g}=\frac{1}{16\pi}\int d^{4} x
\sqrt{-\tilde{g}}\left(\tilde{R}
-2\tilde{g}^{\mu\nu}\partial_{\mu}\phi\partial_{\nu}\phi-4V(\phi)\right),
\ee where $\phi=\sqrt{3}\ln \Phi/2$ and the redefined potential in
the Einstein frame is $V(\phi)=\Phi^{-2}(\phi)U(\Phi(\phi))/4$.

Intervals in Einstein and Jordan frames are linked by the relation
\be \label{metric2} d\tilde{s}^{2}=\Phi
ds^{2}=-e^{2\tilde{\psi}}d{t}^{2}+e^{2\tilde{\phi}}{\tilde{dr}}^{2}+\tilde{r}^2d\Omega^2.
\ee Here we write $d\tilde{s}^{2}$ in form equivalent to
(\ref{metric}) but with different functions $\tilde{\psi}$ and
$\tilde{\lambda}$.

From Eq. (\ref{metric2}) we have that $\tilde{r}^2=\Phi r^{2}$ and
$e^{2\tilde{\psi}}=\Phi e^{2\psi}$. Combining it with equality,
$$
\Phi e^{2\lambda}dr^{2}=e^{2\tilde{\lambda}}d\tilde{r}^{2}
$$
we obtain that,
$$
e^{-2\lambda}=e^{-2\tilde{\lambda}}\left(1-\tilde{r}\phi'(\tilde{r})/\sqrt{3}\right)^{2}.
$$
Gravitational mass $m(r)$ is defined in Jordan frame as, \be m(r)
= \frac{r}{2}\left(1-e^{-2\lambda}\right). \ee We can define
function $\tilde{m}(\tilde{r})$ by the same relation,
$$
\tilde{m}(\tilde{r}) = \frac{\tilde{r}}{2}\left(1-e^{-2\tilde{\lambda}}\right).
$$
Note that $\tilde{m}(\tilde{r})$ is not the gravitational mass
measured by an observer. But the measured gravitational mass
$m(r)$ can be obtained from a simple relation
$\tilde{m}(\tilde{r})$ as \be m(\tilde{r})=
\frac{\tilde{r}}{2}\left(1-\left(1-\frac{2\tilde{m}}{\tilde{r}}\right)\left(1-\tilde{r}\phi'(\tilde{r})/\sqrt{3}\right)^{2}\right)e^{-\phi/\sqrt{3}}
\ee The resulting equations for the metric functions
$\tilde{\lambda}$ and $\tilde{\psi}$ are very similar to the TOV
equations with redefined energy and pressure, and with additional
terms with the energy density and pressure of the scalar field
$\phi$ being: \be \label{TOV1-1} \frac{1}{\tilde{r}^2}\frac{d
\tilde{m}}{d\tilde{r}}=4\pi
e^{-4\phi/\sqrt{3}}\rho+\frac{1}{2}\left(1-\frac{2\tilde{m}}{\tilde{r}}\right)\left(\frac{d\phi}{d\tilde{r}}\right)^{2}+V(\phi),
\ee

\be \label{TOV2-1}
\frac{1}{p+\rho}\frac{dp}{d\tilde{r}}=-\frac{\tilde{m} + 4\pi
e^{-4\phi/\sqrt{3}} p
\tilde{r}^3}{\tilde{r}(\tilde{r}-2\tilde{m})}-
\frac{\tilde{r}}{2}\left(\frac{d\phi}{d\tilde{r}}\right)^{2}+\ee
$$
+\frac{\tilde{r}^2
V(\phi)}{\tilde{r}-2\tilde{m}} +
\frac{1}{\sqrt{3}}\frac{d\phi}{d\tilde{r}} ,
$$

The second
equation is obtained with using condition of hydrostatic
equilibrium,
\begin{equation}\label{hydro-1}
    \frac{dp}{d\tilde{r}}=-(\rho
    +p)\left(\frac{d\psi}{d\tilde{r}}-\frac{1}{\sqrt{3}}\frac{d\phi}{d\tilde{r}}\right).
\end{equation}
Finally, one needs to add the equation of the scalar field
obtained by taking the trace of Einstein equations: \be
\label{TOV3-1} \triangle_{\tilde{r}}
\phi-\frac{dV(\phi)}{d\phi}=-\frac{4\pi}{\sqrt{3}}
e^{-4\phi/\sqrt{3}}(\rho-3p). \ee Here $\triangle_{\tilde{r}}$ is
radial part of Laplace-Beltrami operator for the metric
(\ref{metric2}):
$$
\triangle_{\tilde{r}}=e^{-2\tilde{\lambda}}\left(\frac{2}{r}+\frac{d\psi}{dr}-\frac{d\lambda}{dr}\right)\frac{d}{dr}+e^{-2\tilde{\lambda}}\frac{d^2}{dr^2}.
$$
We rewrite equations (\ref{TOV1-1}), (\ref{TOV2-1}), (\ref{TOV3-1}) in terms of dimensionless variables introduced earlier:
\begin{equation}\label{TOVmod-1}
    \frac{d\tilde{\mu}}{dx} = 4\pi \tilde{x}^2 \theta^{n} \e^{-4\phi/\sqrt{3}} + \frac{\tilde{x}^2}{\beta}\left(\frac{1}{2}\left(1-\frac{2\beta \tilde{\mu}}{\tilde{x}}\right)\left(\frac{d\phi}{d\tilde{x}}\right)^{2}+v(\phi)\right),
\end{equation}
\begin{equation}\label{TOVmod-2}
    \frac{1}{1+4\pi\beta\theta/(n+1)}\frac{d\theta}{d\tilde{x}}=-\frac{1}{4\pi}\frac{\tilde{\mu}+16\pi^{2}\beta\theta^{n+1}\tilde{x}^3 e^{-4\phi/\sqrt{3}}}{\tilde{x}(\tilde{x}-2{\beta}\tilde{\mu})}-
\end{equation}
$$
-\frac{\tilde{x}^2}{4\pi \beta(\tilde{x}-2\beta\tilde{\mu})}\left(\frac{1}{2}\left(1-\frac{2\beta \tilde{\mu}}{\tilde{x}}\right)\left(\frac{d\phi}{d\tilde{x}}\right)^{2}-v(\phi)\right)+\frac{1}{4\sqrt{3}\pi\beta}\frac{d\phi}{d\tilde{x}}
$$
Here we introduced the dimensionless potential of scalar field $v(\phi)$,
$$
v(\phi)=a^2 V(\phi).
$$
The equation for the scalar field $\phi$ after some calculations
can be written in the following form,
\begin{equation}\label{EqSc}
    \left(1-\frac{2\beta \tilde{\mu}}{\tilde{x}}\right)\left(\frac{d^2 \phi}{d\tilde{x}^2}+\left(\frac{2}{\tilde{x}}-\frac{4\pi\beta}{1+4\pi\beta/(n+1)}\frac{d\theta}{d\tilde{x}}+\frac{1}{\sqrt{3}}\frac{d\phi}{d\tilde{x}}\right)\frac{d\phi}{d\tilde{x}}\right)+
\end{equation}
$$
+\left(\frac{\beta \tilde{\mu}}{\tilde{x}^2}-\frac{\beta}{\tilde{x}}\frac{d\tilde{\mu}}{d\tilde{x}}\right)\frac{d\phi}{d\tilde{x}}-\frac{dv}{d\phi} = 
$$
$$
=-\frac{4\pi\beta}{\sqrt{3}}e^{-4\phi/\sqrt{3}}\theta^{n}\left(1-12\pi\beta\theta/(n+1)\right).
$$

Equations (\ref{TOVmod-1}), (\ref{TOVmod-2}) with (\ref{EqSc}) can
be integrated numerically for various values of the parameter $n$.
From previous analysis of TOV equations in a case of white dwarfs
we know that terms proportional to small parameter $\beta$ do not
affect considerably the solution. Assuming the same in a case of
modified gravity we can study  the``reduced'' system of equations,
leaving only terms with scalar field in which parameter $\beta$ in
denominator:
\begin{equation}\label{TOVred-1}
    \frac{d\tilde{\mu}}{dx} = 4\pi \tilde{x}^2 \theta^{n} \e^{-4\phi/\sqrt{3}} + \frac{\tilde{x}^2}{\beta}\left(\frac{1}{2}\left(\frac{d\phi}{d\tilde{x}}\right)^{2}+v(\phi)\right),
\end{equation}
\begin{equation}\label{TOVred-2}
    \frac{d\theta}{d\tilde{x}}=-\frac{\tilde{\mu}}{4\pi \tilde{x}^{2}}-\frac{\tilde{x}}{4\pi \beta}\left(\frac{1}{2}\left(\frac{d\phi}{d\tilde{x}}\right)^{2}-v(\phi)\right)+
    \frac{1}{4\sqrt{3}\pi\beta}\frac{d\phi}{d\tilde{x}}.
\end{equation}
In the left hand side of Eq. (\ref{EqSc}) we also drop terms
containing parameter $\beta$ and terms with square of first
derivative of scalar field. In the right hand side of this
equation we leave only terms with first power of $\beta$:
\begin{equation}\label{Eqred}
    \frac{d^2 \phi}{d\tilde{x}^2}+\frac{2}{\tilde{x}}\frac{d\phi}{d\tilde{x}}-\frac{dv}{d\phi}=-\frac{4\pi\beta}{\sqrt{3}}e^{-4\phi/\sqrt{3}}\theta^{n}.
\end{equation}
The system of equations should be complemented by initial
conditions at the center of star:
$$
\theta(0)=1,\quad \tilde{\mu}(0)=0, \quad \phi(0)=\phi_{0}, \quad \frac{d\phi(0)}{d\tilde{x}}=0.
$$
The condition of asymptotic flatness requires that
$$\phi\rightarrow 0 \quad \mbox{at} \quad x\rightarrow \infty. $$
It is convenient to analyse system of equations in the Einstein
frame and then after calculations go back to the Jordan frame. We
are interested mainly in effects of modified gravity and therefore
we consider reduced system of equations (\ref{TOVred-1}),
(\ref{TOVred-2}), (\ref{Eqred}).


\section{Simple model of R2 gravity: perturbative approach and numerical integration of reduced system}

Considering a simple $R^2$ gravity with $f(R)=R+\alpha R^2$
gravity we have that,
\begin{equation}\label{simple}
    v(\phi)=\frac{1}{16\tilde{\alpha}}\left(1-e^{-2\phi/\sqrt{3}}\right)^{2},\quad \tilde{\alpha}=\alpha/a^2.
\end{equation}
Usually one assumes that $\alpha>0$, otherwise model of $R^2$
gravity leads to instabilities. Because scalar field is very small
we can expand potential $v(\phi)$ leaving only first non-zero
term:
$$
v(\phi)=\frac{1}{12\tilde{\alpha}}\phi^2.
$$
Assuming that the scalar potential term is dominant and scalar
field is very small we can reduce equation (\ref{Eqred}) to
relation between the density and the scalar field:
\begin{equation}\label{EqScapp}
\frac{\phi}{6\tilde{\alpha}}=\frac{4\pi\beta}{\sqrt{3}}\theta^{n}.
\end{equation}
From this approximation follows that outside the star $\phi = 0$
and $\frac{d\phi}{dx} = 0$. Therefore, $\tilde{x}_{f} = x_{f}$ and
$\tilde{\mu}(\tilde{x}_{f}) = {\mu}(x_{f})$.

From observations it follows that the upper limit on parameter
$\alpha$ is $\sim 10^{15}$ cm$^{2}$. For $\alpha=10^{14}$ cm$^{2}$
the results of the calculations are given in Table I.

The analysis shows that the contribution of the scalar field on
the pressure and the density is very negligible. Only the last
term in (\ref{TOVred-2}) gives a considerable effect on the
solution of  the equations. It is very easy to understand why this
happens. From the approximation (\ref{EqScapp}), it follows that
$d\phi/dx \sim \alpha \mathcal{O}(\beta)$ and therefore $$
\frac{1}{2}\left(\frac{d\phi}{dx}\right)^{2} \sim
10^{2}\tilde{\alpha}^2 \mathcal{O}(\beta^2), \quad v(\phi)\sim
10\tilde{\alpha} \mathcal{O}(\beta^2).
$$
The length parameter $a$ varies from $2.25\times 10^{8}$ cm for
$\rho_c = 10^{7}$ g/cm$^3$ to $4.85\times 10^{7}$ cm for $\rho_c =
10^{9}$ g/cm$^3$.  Therefore, even for the upper limit of the
parameter $\alpha$, the dimensionless parameter $\tilde{\alpha}<1$
and the contribution of the terms in brackets in (\ref{TOVred-1}),
(\ref{TOVred-2}) is $\mathcal{O}(\beta)$ i.e. it is comparable
with relativistic effects from General Relativity in comparison
with Newtonian gravity.

Only for sufficiently large $\tilde{\alpha}$ one can expect that
the approximation (\ref{EqScapp}) does not work because the square
of scalar field derivative is comparable to the value with
potential term.

We investigate solutions of the reduced system of equations
without approximation (\ref{EqScapp}) and found that account of
more exact solution for scalar field increases mass of star. For
$\tilde{\alpha}<0.02$ this increase is less than $0.01 M_{\odot}$
in comparison to the perturbative solution. In the case of
$\alpha=10^{14}$ cm$^2$ this corresponds to central densities up
to $3.2\times 10^8$ g/cm$^3$. For $\alpha=10^{13}$ cm$^2$, the
approximation (\ref{EqScapp}) can be used for central densities
$\rho_c<10^{10}$ g/cm$^3$. In real white dwarfs as assumed central
densities are less. For $\alpha=10^{15}$ cm$^2$ one needs to solve
the exact equation for the scalar field. Of course such values of
the parameter $\alpha$ represent only theoretical interest because
the decrease of white dwarf mass is very large in comparison with
GR which is difficult to reconcile with the available
observational data.

The profile of the scalar field from solution of (\ref{Eqred}) is
a function decreasing with the coordinate $x$ and follows to the
density profile. For illustration we plot the solution of
(\ref{Eqred}) and profile of scalar field derived from
(\ref{EqScapp}) for various values of $\rho_c$ and
$\alpha=10^{14}$ cm$^2$. For large $\tilde{\alpha}$, the profile
of the exact solution differs significantly from the
approximation. We also see that the ``tale'' of the scalar field
outside the surface of the star is very short and this existence
does not affect the stellar mass. We point out that another
situation takes place in neutron stars. Density sharply drops near
the surface of NS, but the scalar field decreases more slowly and
therefore in $R^2$ gravity around surface of neutron star, a
``gravitational sphere'' exists with scalar curvature $R\neq 0$
(or $\phi\neq 0$ in Einstein frame). It gives contribution to
gravitational mass and for high central densities and $\alpha>0$
NS mass increases.
\begin{figure}
    \centering
    \includegraphics[scale=0.35]{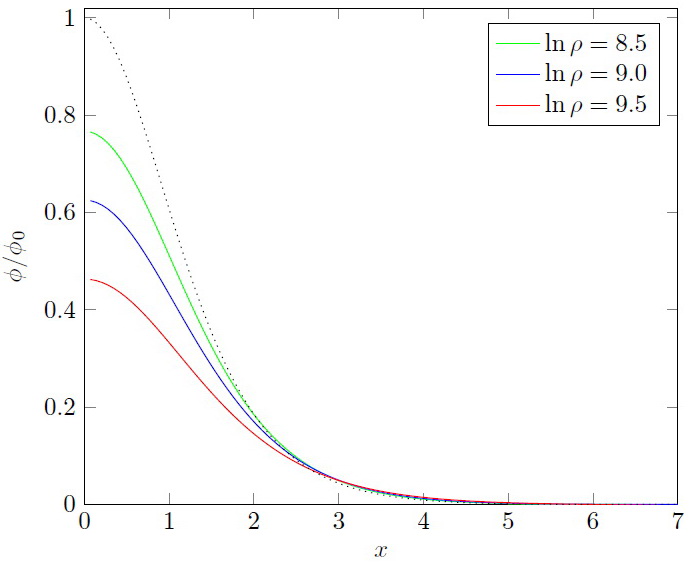}
    \caption{Profile of scalar field (solid lines) as function of dimensionless variable x in comparison with approximation (\ref{EqScapp}) (black dotted line) for some  central densities. Parameter $\alpha=10^{14}$ cm$^2$. $\phi_0$ means value of scalar field in the center of star for perturbative solution. For exact solution scalar field starts from smaller values ($\phi(0)/\phi_0<1$).}
    \label{fig:1}
\end{figure}
\begin{table}
\begin{center}
\begin{tabular}{|c|c|c|c|c|}
\hline
$\ln{\rho_c}$ & $\tilde{\alpha},$ & $\phi_{c}/\phi_{p}(0)$ & $M,$ & $M_{p},$ \\
& $10^{-3}$ &  & $M_{\odot}$ & $M_{\odot}$ \\
\hline
7 & 1.97 & $\sim 1$  & 1.448 & 1.448   \\
\hline
7.5 & 4.25 & $\sim 1$ & 1.439 & 1.439  \\
\hline
8 & 9.15 & 0.92203 & 1.419 & 1.418  \\
\hline
8.5 & 19.71 & 0.76707 & 1.389 & 1.377   \\
\hline
9 & 42.48 & 0.62547 & 1.334 &  1.295 \\
\hline
9.5 & 91.52 & 0.46316 & 1.260 & 1.162  \\
\hline
\end{tabular}
\caption{\label{tab:singularity1} Difference between results for
stellar masses from exact solution of reduced system ($M$) and
perturbative solution ($M_p$) for $\alpha=10^{14}$ cm$^2$. We give
also corresponding values of dimensionless parameter
$\tilde{\alpha}$ and the relation $\phi_{c}/\phi_{p}(0)$. Here
$\phi_c$ is the value of scalar field in the center of star and
$\phi_p(0)$ is the value of scalar field from perturbative
approximation.}
\end{center}
\end{table}

\section{Realistic equation of state}

The next step is to consider more realistic EoSs. We choose the
Chandrasekhar EoS for stellar matter, which can be written in
parametric form:
\begin{equation}\label{ChEoS}
\rho = B y^3,
\end{equation}
$$
p=A \left[y(2y^2-3)(1+y^2)^{1/2}+3\ln(x+(1+x^2)^{1/2}\right],
$$
$$
B = 9.82\times 10^{5} \mu_{e}\mbox{ g/cm}^{3}, A = 6.02\times 10^{22}\mbox{ dyne/cm}^{2}.
$$
Again, it is useful to introduce dimensionless variables in Eqs.
(\ref{TOV1-1}), (\ref{TOV2-1}). Taking into account characteristic
radii and masses of white dwarfs let us define:
$$
\tilde{r} = R_{e} \tilde{x}, \quad \tilde{m} = M_\odot \tilde{\mu},
$$
$$
\rho = \rho_c \eta, \quad p = p_{c} \xi.
$$
Here $R_{e}$ means radius of Earth. Restoring $G$ and $c$ in
equations we obtain the following system of equations for
dimensionless variables $\tilde{\mu}$, $\eta$, $\xi$ and $\phi$:
\begin{equation}
\frac{d\tilde{\mu}}{d\tilde{x}} =\delta  4\pi \tilde{x}^{2} e^{-4\phi/\sqrt{3}}\eta + \frac{1}{2}\frac{R_{e}}{r_{g}}\left(1-\frac{2r_{g}}{R_{e}}\frac{\tilde{\mu}}{\tilde{x}}\right)\left(\frac{d\phi}{d\tilde{x}}\right)^{2} +\frac{R_{e}}{r_g}v(\phi),
\end{equation}
\begin{equation}
    \frac{1}{\eta+\xi/\gamma}\frac{d\xi}{d\tilde{x}}=-\gamma\frac{r_{g}}{R_{e}}\frac{\tilde{\mu}}{\tilde{x}(\tilde{x}-2r_{g}\tilde{\mu}/R_{e})}-\delta\frac{r_{g}}{R_{e}}\frac{4\pi e^{-4\phi/\sqrt{3}}\tilde{x}^{3}\xi}{\tilde{x}(\tilde{x}-2r_{g}\tilde{\mu}/R_{e})}+
\end{equation}
$$
+ \gamma\left(-\frac{\tilde{x}}{2}\left(\frac{d\phi}{d\tilde{x}}\right)^{2}+\frac{\tilde{x}^{2} v(\phi)}{\tilde{x}-2r_{g}\tilde{\mu}/R_{e}} + \frac{1}{\sqrt{3}}\frac{d\phi}{d\tilde{x}}\right),
$$
\begin{equation}
    \left(1-\frac{2r_{g}}{R_{e}}\frac{\tilde{\mu}}{\tilde{x}}\right)\left(\frac{d^{2}\phi}{d\tilde{x}^{2}}+\left(\frac{2}{\tilde{x}}-\frac{1}{\gamma\eta+\xi}\frac{d\xi}{dx}+\frac{1}{\sqrt{3}}\frac{d\phi}{d\tilde{x}}\right)\frac{d\phi}{d\tilde{x}}\right)+
\end{equation}
$$
+\frac{r_{g}}{R_{e}}\left(\frac{\tilde{\mu}}{\tilde{x}^2}-\frac{1}{\tilde{x}}\frac{d\tilde{\mu}}{d\tilde{x}}\right)\frac{d\phi}{d\tilde{x}}-\frac{dv}{d\phi} = -\frac{4\pi\delta}{\sqrt{3}}\frac{r_{g}}{R_{e}}e^{-4\phi/\sqrt{3}}\left(\eta-3\xi/\gamma\right),
$$
where the dimensionless parameters $\delta$ and $\gamma$ are introduced:
$$
\delta = \frac{\rho_c R_{e}^{3}}{M_{\odot}},\quad \gamma = \frac{\rho_c c^2}{p_c}
$$
and $v(\phi)=V(\phi) R_{e}^{2}$. The parameter $r_{g} =
{GM_{\odot}}/{c^2}$ is nothing else than half of gravitational
radius of Sun. Again we consider the reduced system of equations
neglecting terms containing the relation $r_{g}/R_{e}<<1$ in the
denominators and take into account that that the parameter
$\gamma>>\delta$ for white dwarfs, and we get,
\begin{equation}\label{eq30}
\frac{d\tilde{\mu}}{d\tilde{x}} =\delta  4\pi \tilde{x}^{2} e^{-4\phi/\sqrt{3}}\eta + \frac{R_e}{r_g}\left(\left(\frac{1}{2}\frac{d\phi}{d\tilde{x}}\right)^{2}+v(\phi)\right),
\end{equation}
\begin{equation}\label{eq31}
    \frac{1}{\eta}\frac{d\xi}{d\tilde{x}}=-\gamma\frac{r_{g}}{R_{e}}\frac{\tilde{\mu}}{\tilde{x}^{2}}+ \gamma\left(-\frac{\tilde{x}}{2}\left(\frac{d\phi}{d\tilde{x}}\right)^{2}+{\tilde{x} v(\phi)} + \frac{1}{\sqrt{3}}\frac{d\phi}{d\tilde{x}}\right),
\end{equation}
\begin{equation}\label{eq32}
   \frac{d^{2}\phi}{d\tilde{x}^{2}}+\left(\frac{2}{\tilde{x}}+\frac{1}{\sqrt{3}}\frac{d\phi}{d\tilde{x}}\right)\frac{d\phi}{d\tilde{x}}-\frac{dv}{d\phi} = -\frac{4\pi\delta}{\sqrt{3}}\frac{r_{g}}{R_{e}}e^{-4\phi/\sqrt{3}}\eta.
\end{equation}
As in the previous case we consider the potential for $R^2$
gravity and compare results from perturbative approximation for
the scalar field and more exact solution of (\ref{eq32}). The main
result is the same as for relativistic polytrope: the stellar mass
decreases in comparison with GR for the same central density. For
large densities one needs to solve the equation for scalar field
because the perturbative approximation is not valid. But it is
interesting to note that in the case of $R^2$ gravity, the stellar
mass has a maximum for some central density and then the mass
decreases although in GR mass grows with density for the
Chandrasekhar EoS.

In fig. 2 we depicted the mass-density relation for the interval
of densities between $10^{7}$ and $10^{10}$ g/cm$^3$ for various
values of $\alpha$. These results are important for establishing
the upper limit on parameter $\alpha$ in $R^2$ gravity. According
to latest observations white dwarfs with masses $M>1.3 M_{\odot}$
are very rare. Only 25 such white dwarfs are observed in vicinity
of Sun System (\citealp{Cilic}). White dwarf J1329 + 2549 is currently
the most massive white dwarf known with a mass of $1.351\pm 0.006
M_{\odot}$. Considering $1.35M_\odot$ as lower limit on maximal
value for white dwarf mass and assuming that the Chandrasekhar EoS
is valid, we conclude that $\alpha<10^{13}$ cm$^2$. These results
lead to conclusion that GR gives satisfactory picture for white
dwarfs parameters. Analyzing of white dwarfs parameters in $R^2$
gravity for realistic values of $\alpha$ can be performed using
approximation for scalar field.
\begin{figure}
    \centering
    \includegraphics[scale=0.35]{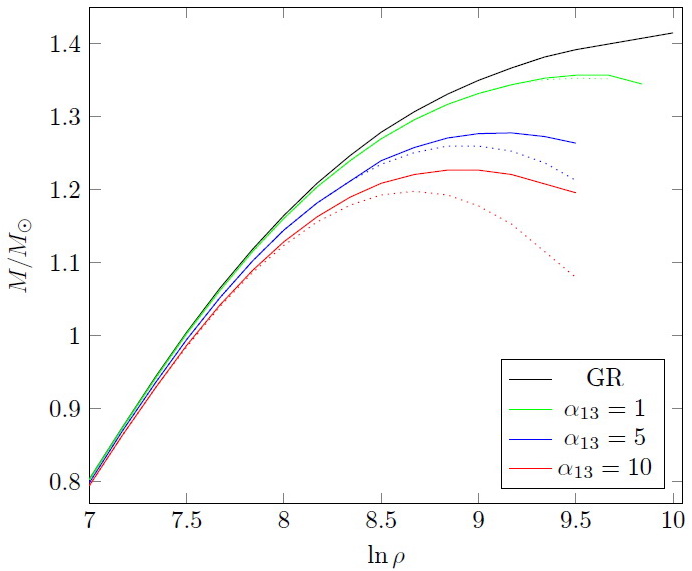}
    \caption{Mass-density relation in $R^2$ gravity for some $\alpha$ in
comparison with GR. The dotted lines correspond to results
obtained with simple approximation for scalar field. $\alpha_{13}$
means that value of $\alpha$ is given in units of $10^{13}$
cm$^2$.}
    \label{fig:2}
\end{figure}

\begin{figure}
    \centering
    \includegraphics[scale=0.35]{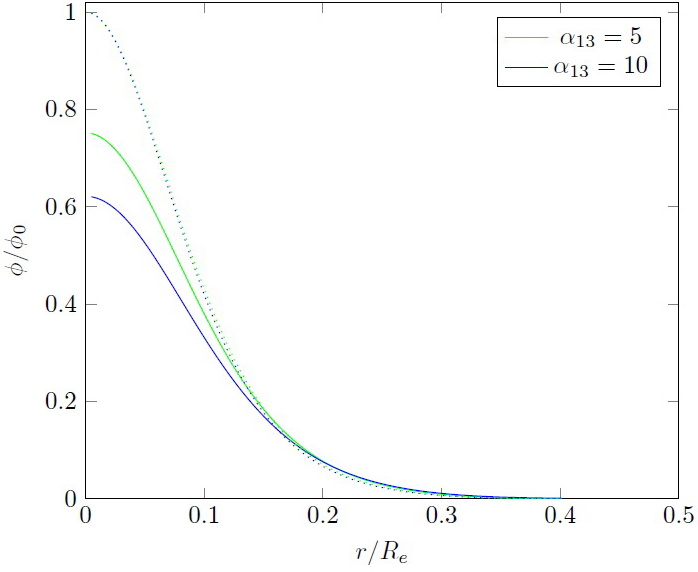}
    \caption{Profile of the scalar field (solid lines) as function of radial
coordinate in comparison with approximation (\ref{EqScapp})
(dotted lines) for $\rho_c=10^9$ g/cm$^3$ and Chandrasekhar EoS.}
    \label{fig:3}
\end{figure}
From our results follows that for $\rho_c>\rho_c^{crit}$, white
dwarfs are unstable in $R^2$ gravity. Critical density and minimal
radius of white dwarf depends on the value of $\alpha$. In light
of these results for masses and radii of white dwarfs near the
Chandrasekhar limit, in principle they allow us to estimate the
upper limit of $\alpha$ more precisely.

The scalar field obtained from the numerical solution of
Eqs.(\ref{eq30}), (\ref{eq31}), (\ref{eq32}) decreases from the center
to the surface of the star in the same manner as for the case of
polytropic EoS: it starts from $\phi(0)<\phi_{0}$ where $\phi_0$
is the central value of the perturbative solution and then follows
the density profile (see Fig. 3).

\section{Chandrasekhar limit of mass in another models of modified gravity}

We showed that in the case of white dwarfs in $R^2$ gravity for
realistic parameters we can neglect the derivatives of the scalar
field in its field equation and use simple approximation for the
scalar field. As we showed , the profile of scalar field is a
monotonic function of the radial coordinate. It is interesting to
investigate another model of modified gravity.

Let's consider model with $f(R)=R+\alpha^{l-1} R^{l}$ where $l >
2$. This representation is chosen so that parameter $\alpha$ has a
dimension of the square of length. The potential of the scalar
field theory in the corresponding equivalent scalar-tensor theory
in this case is \be V(\Phi)=
D\Phi^{-2}\left(\Phi-1\right)^{\frac{l}{l-1}}, \quad
D=\frac{l-1}{4l^{\frac{l}{l-1}}}\alpha^{-1}, \quad
\Phi=e^{2\phi/\sqrt{3}}. \ee Again if the scalar field $\phi$ is
very small, one can expand the expression for $V(\phi)$ and
obtain,
$$
V(\phi)\approx D \left(\frac{2}{\sqrt{3}}\right)^{\frac{l}{l-1}}\phi^{\frac{l}{l-1}}.
$$
If the potential term dominates we can use approximation,
\begin{equation}
    \phi \approx \frac{l(2\sqrt{3})^{l}}{4}\left(\frac{\alpha}{a^2}\right)^{l-1}\left(\frac{4\pi \beta}{\sqrt{3}}\right)^{l-1}\theta^{n(l-1)}.
\end{equation}
The dimensionless potential and the square of the scalar field
derivative in order of magnitude are,
$$
v(\phi) \sim \left(\frac{\alpha}{a^{2}}\right)^{l-1}\mathcal{O}(\beta^{l}),\quad  \left(\frac{d\phi}{dx}\right)^{2} \sim \left(\frac{\alpha}{a^{2}}\right)^{2(l-1)}\mathcal{O}(\beta^{2l-2})
$$
One can expect that for realistic values of $\alpha/a^{2}<<1$
approximation for the scalar field is valid. And as in the case of
$R^2$ gravity, the effects of the scalar field on the density and
the pressure are negligible. Moreover, for $m \geq 3$ the effects
of modified gravity will be of the next order of smallness on the
parameter $\beta$ in comparison  with the relativistic effects of
GR on background of Newton's gravity. The square of the scalar
field derivative is an order lower in comparison with potential
term for $l>2$.

Calculations for some $l>2$ show the same pattern as for $l=2$:
The stellar mass decreases with increasing central density. Some
results are given in Fig. 4. For $\alpha/a^2 \sim \mathcal{O}(1)$
perturbative solution is not valid. Obviously, that for realistic
Chandrasekhar EoS we obtain that stellar mass has a maximum for
certain density as in the case of $R^2$ gravity.
\begin{figure}
    \centering
    \includegraphics[scale=0.35]{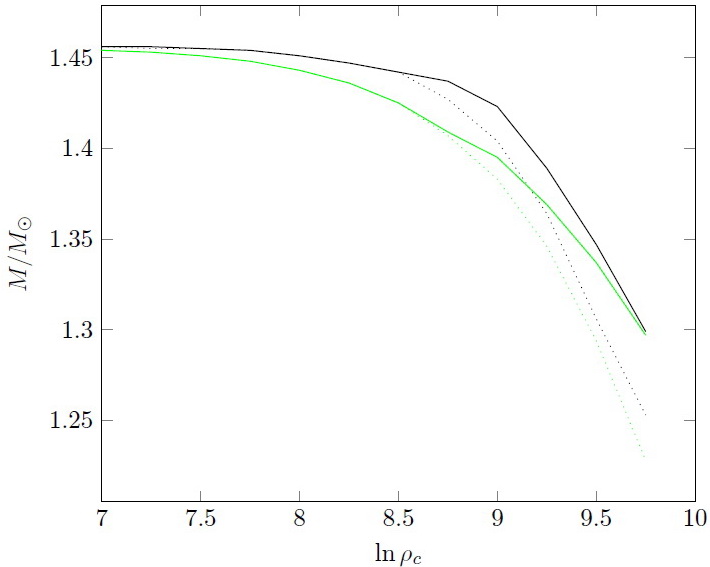}
    \caption{Mass-density relation for polytrope with $n=3$ in
$R+\alpha^{l-1}R^{l}$ gravity for ($\alpha=10^{14}$ cm$^2$,
$l=2.1$, green lines) and ($\alpha=5\times 10^{14}$ cm$^2$,
$l=2.4$, black lines). The dotted lines correspond to results
obtained with simple approximation for scalar field.}
    \label{fig:4}
\end{figure}
If we assume that the perturbative approximation for the scalar
field is valid, then the scalar field is a monotonic function of
$\phi$ because $dv/d\phi>0$ as follows from relation
\begin{equation}\label{approx}
\frac{dv}{d\phi} = \frac{4\pi\beta}{\sqrt{3}}\theta^{n}.
\end{equation}
Of course $\theta$ is a monotonically decreasing function of
coordinate for stable stellar configurations. Therefore, for the
scalar field we can write that,
$$
\phi = F(\theta)
$$
where $F$ is a monotonic increasing function of its argument. The
scalar field decreases with the coordinate and the potential tends
to its minimum on the surface of star. The following conditions
should be satisfied: \be F(\theta)=0, \quad \frac{dF}{d\theta} =
0\mbox{ for } \theta=0. \ee These conditions guarantee that scalar
field and its first derivative outside of star vanish. Of course
for many potentials of scalar field, the explicit relation between
the scalar field and $\theta$ does not exist. The first derivative
of the scalar field with respect to the coordinate $\tilde{x}$ is,
$$
\frac{d\phi}{d\tilde{x}} = \frac{dF}{d\theta}\frac{d\theta}{d\tilde{x}}.
$$
Equation (\ref{TOVred-2}) by taking into account the expression
for the derivative of the scalar field and without the term
$(d\phi/dx)^{2}$ can be written as,
\begin{equation}
\left\{1-\frac{1}{4\sqrt{3}\pi\beta}\frac{dF}{d\theta}\right\}\frac{d\theta}{dx}=-\frac{\tilde{\mu}}{4\pi\tilde{x}^2}+\frac{\tilde{x}}{4\pi\beta}v(\phi).
\end{equation}
Then after some simple algebra, one obtains the analog of the
Lane-Emden equation:
\begin{equation}\label{LE-eq}
   \frac{d}{d\tilde{x}}\left\{\tilde{x}^2\left\{1-\frac{1}{4\sqrt{3}\pi\beta}\frac{dF}{d\theta}\right\}\frac{d\theta}{dx}\right\} = - \tilde{x}^{2}\theta^{n} e^{-4 \phi /\sqrt{3}}+
\end{equation}
$$
+\frac{\tilde{x}^{2}}{2\pi \beta} v(\phi) + \frac{\tilde{x}^{3}}{\sqrt{3}}\theta^{n}\frac{dF}{d\theta}\frac{d\theta}{d\tilde{x}}.
$$
To obtain the scalar field potential as function of $\theta$ in
frames of our approximation, one needs to take the following
integral:
\begin{equation}\label{Vapprox}
v(\theta) = \frac{4\pi \beta}{\sqrt{3}}\int_{0}^{\theta} \frac{dF}{d\theta}\theta^{n} d\theta.
\end{equation}
Therefore for given $F(\theta)$ we have the potential of the
scalar field in parametric form. Of course the explicit form of
$v(\phi)$ can be written only for relatively simple functions
$F(\theta)$. The function $\tilde{\mu}$ is defined from the
following relation: \be \tilde{\mu}(x) = -4\pi
\tilde{x}^{2}\frac{d\theta(\tilde{x})}{d\tilde{x}}+\frac{\tilde{x}^{3}}{\beta}v(\theta)+\frac{\tilde{x}^{2}}{\sqrt{3}\beta}\frac{dF}{d\theta}\frac{d\theta(\tilde{x})}{d
\tilde{x}}. \ee Realistic solutions of Eq.(\ref{LE-eq}) for chosen
$F(\theta)$ posses the same property as the solution of the
Lane-Emden equation in Newtonian gravity: for some $\tilde{x}_{f}$
the function $\theta$ vanishes. This $\tilde{x}_{f}$ corresponds
to surface of white dwarf. In Jordan frame $x_{f}=\tilde{x}_{f}$
because scalar field $\phi$ is zero on the star's surface and
therefore $\Phi = 1$. Because the first derivative of the scalar
field also vanishes on the surface, the gravitational mass of
white dwarf is,
$$
\mu = \tilde{\mu}(\tilde{x}_{f})
$$
and therefore,
$$
\mu = -4\pi \tilde{x}^{2}_{f}\frac{d\theta(\tilde{x}_{f})}{d\tilde{x}}
$$
because $v(\theta)=0$ and $dF/d\theta=0$ for $\theta=0$.

Because the curvature $R$ in the case of white dwarfs is
relatively small, one can propose that a realistic model of $f(R)$
gravity function $f(R)$ can be represented as series in powers of
$R$:
$$
f(R)=R + \alpha_{1} R^{2} + \alpha_{2}^{2} R^{3} + ...
$$
For $R\rightarrow 0$ and $\alpha_{1}\neq 0$, the corresponding
potential of the equivalent scalar-tensor theory is,
$$
v(\phi)\rightarrow \frac{\phi^{2}}{12\alpha_{1}}.
$$
If $\alpha_2 \neq 0$ and $\alpha_{1}=0$ the potential $v(\phi)\sim
\phi^{3/2}$ and so on. Therefore, the first derivative of
potential $dv/d\phi$ should contain terms $\sim \phi^{1/(k-1)}$,
k=2,3... and monotonic function $F(\theta)$ is a sum
$$
F(\theta) = s_{1}\theta^{n} + s_{2}\theta^{2n}+...
$$
on interval $0\leq \theta \leq 1$. From previous results we
conclude that stellar mass decreases for this function $F(\theta)$
with increasing of central density. Therefore, in frames of
perturbative approach for realistic $f(R)$, one should expect that
the mass of the white dwarf decreases in comparison with the GR
case for the same central density.

If the function $F(\theta)$ decreases with its argument this
corresponds to increasing scalar field $\phi$ from center to
surface. If the scalar field is defined from (\ref{approx}) this
leads to an increase of the potential term from the center to the
surface. For $R^2$ gravity such situation takes place for
$\alpha<0$. But this model of gravity cannot be considered as
realistic because it leads to instabilities.

Therefore we conclude that in frames of perturbative approach it
is impossible to construct solutions of the system
(\ref{TOVred-1}), (\ref{TOVred-2}), (\ref{Eqred}) such that
gravitational mass increases in comparison with GR. Increase of
the value takes place only for unrealistic models of $F(R)$
gravity.

Non-monotonic function $F(\theta)$ as follows from (\ref{Vapprox})
leads to that potential of scalar field should be an ambiguous
function of its argument. Does a solution of (\ref{Eqred}) exists
such that the derivative of the scalar field changes the sign? Let
us assume that this indeed happens once. The scalar field starts
from some positive value from the center of the star and reaches a
minimum $\phi_{min}<0$ at some point $x_s$. At the vicinity of the
minimum $d^2\phi/dx^2\geq 0$ and therefore $dv/d\phi>0$ the
potential goes down.  Then the scalar field increases and
asymptotically tends to zero for large x. The asymptotical value
of the scalar field outside the star should correspond to the
minimum of the potential. We have therefore the situation which
for example can take place for potentials of the form $u\sim
\phi^{n}$ where $n=2k$, k=1,2... Scalar field reaches the minimum
for some $x<x_s$ and develops negative values and finally
approaches zero and again the minimum of the potential. But our
consideration shows that for such potentials, we can construct
solutions when scalar field is monotonic decreasing function from
center to surface.

\section{Concluding Remarks}

We investigated the question about the maximal white dwarf mass
limit in $f(R)$ gravity. Our analysis involved a polytropic EoS
with $n=3$ and more realistic Chandrasekhar EoS. Also the
equivalent scalar-tensor theory in the Einstein frame was used
with the subsequent transition to the Jordan picture. For $f(R)$
gravity, one can consider the reduced system of equations, because
relativistic effects of GR in the case of white dwarfs are
negligible in Newtonian gravity background. In models with
$f(R)=R+\alpha^{l-1} R^{l}$ for any $l\geq 2$, the mass of white
dwarfs decreases in comparison with GR for $\alpha^{l-1}>0$. For
realistic values of $\alpha$, the perturbative approach is valid.
It is sufficient to account only potential terms in the equation
for the scalar field and obtain relation for its field. For stable
stars, the density should decrease from center to surface and the
corresponding profile of scalar field also decreases. It is
important to note that the contribution of the scalar field to
energy density is around $\mathcal{O}(\beta^{l-1})$ where
$\beta=\rho_c a^2 G/c^2$ is small relativistic parameter. This
contribution is comparable (for $R^2$ gravity) with the effect
from the relativistic corrections to solutions of Lane-Emden
equation or even less (for $l>2$). Applicability of the
perturbative approach is defined by the relation
$(\alpha/a^2)^{l-1}$. More precise calculations show that the
scalar field starts from some value $\kappa \phi_{p}(0)$ at the
center of star where $0<\kappa<1$ and $\phi_p(0)$ is central value
of the scalar field from approximation. In the case of the
Chandrasekhar EoS, there exists a critical value of the central
density for which the stellar mass reaches a maximum value and
then decreases. Precise estimations of the maximal value of white
dwarfs mass from astronomical observations has significance
towards constraining the upper limit of parameter $\alpha$. If the
Chandrasekhar EoS is valid, we can reconcile observational data
for white dwarfs in $R^2$ gravity only for $\alpha < 10^{13}$
cm$^2$. In comparison with NSs, it is worth to note that as
believed EoS is known much more accurately. Therefore, one can
hope that the possible effects of modified gravity will not
disguise by uncertainty in knowledge of the equation of state. For
NSs also the solution of scalar field has the following feature
namely around of the star area with $\phi\neq 0$ exists. This area
gives a contribution to the gravitational mass and the net effect
for neutron mass with masses $M>1.5M_{\odot}$ is increasing of
mass. For white dwarfs there are no significant ``scalar tails''
because near the surface the perturbative solution is valid with
high accuracy and therefore the scalar field is defined by density
mainly. In the Einstein frame it means that scalar curvature near
the surface is close to its value in GR namely $R\approx 8\pi
(\rho - 3p)$ and drops to zero outside the star very quickly.

\section*{Acknowledgments}

This work was supported by MINECO (Spain), project
PID2019-104397GB-I00 (S.D.O). This work by S.D.O was also
partially supported by the program Unidad de Excelencia Maria de
Maeztu CEX2020-001058-M, Spain. This work was supported by
Ministry of Education and Science (Russia), project
075-02-2021-1748 (AVA).

\label{lastpage}


\begin{thebibliography}{99}

\bibitem[\protect\citeauthoryear{Arapoglu, Deliduman \& Eksi}{2011}]{Arapoglu:2010rz}
Arapoglu A.S., Deliduman C., Eksi K.Y., 2011,
JCAP, {07}, 020 
[arXiv:1003.3179 [gr-qc]]

\bibitem[\protect\citeauthoryear{Astashenok et al.}{2020}]{Astashenok:2020qds}
Astashenok A.V., Capozziello S., Odintsov S.D.,
Oikonomou V.K., 2020,
Phys. Lett. B, {811}, 135910
[arXiv:2008.10884 [gr-qc]]

\bibitem[\protect\citeauthoryear{Astashenok et al.}{2021}]{Astashenok:2021peo}
Astashenok A.V., Capozziello S., Odintsov S.D.,
Oikonomou V.K., 2021,
 Phys. Lett.  B, {816}, 136222
 [arXiv:2103.04144 [gr-qc]]

\bibitem[\protect\citeauthoryear{Astashenok, Capozziello \& Odintsov}{2015}]{Astashenok:2014nua}
Astashenok A.V., Capozziello S., Odintsov S.D., 2015,
JCAP, 01, 001 
[arXiv:1408.3856 [gr-qc]]

\bibitem[\protect\citeauthoryear{Astashenok \& Odintsov}{2020}]{Astashenok:2020cfv}
Astashenok A.V., Odintsov S.D., 2020,
MNRAS, {493}, 78
[arXiv:2001.08504 [gr-qc]]

\bibitem[\protect\citeauthoryear{Astashenok, Odintsov \& Cruz-Dombriz}{2017}]{Alvaro} Astashenok A.V., Odintsov S.D., de la Cruz-Dombriz A., 2017, Class. Quant. Grav., 34, 205008 [arXiv:1704.08311 [gr-qc]]

\bibitem[\protect\citeauthoryear{Bl\'azquez-Salcedo, Scen Khoo \& Kunz}{2020}]{Blazquez-Salcedo:2020ibb}
Bl\'azquez-Salcedo J.L., F.~Scen Khoo and J.~Kunz,
EPL \textbf{130} (2020) no.5, 50002
[arXiv:2001.09117 [gr-qc]].

\bibitem[\protect\citeauthoryear{Capozziello \& Laurentis}{2011}]{reviews1}
 Capozziello S., De Laurentis M., 2011,
   Phys.\ Rept.,  {509}, 167
   [arXiv:1108.6266 [gr-qc]].

\bibitem[\protect\citeauthoryear{Capozziello \& Faraoni}{2011}]{reviews2}
Capozziello S., Faraoni V.
 \textit{Beyond Einstein Gravity : A Survey of Gravitational Theories for Cosmology and Astrophysics}, 2011,
  Fundam.\ Theor.\ Phys.,  {170}, Springer, Dordrecht

\bibitem[\protect\citeauthoryear{Capozziello et al.}{2016}]{Capozziello:2015yza}
Capozziello S.,, De Laurentis M., Farinelli R., Odintsov S.D., 2016,
Phys. Rev. D, {93}, 023501
[arXiv:1509.04163 [gr-qc]]

\bibitem[\protect\citeauthoryear{Chew et al.}{2019}]{Chew:2019lsa}
Chew X.Y., Kleihaus B.,
Kunz J., Dzhunushaliev V., Folomeev V., 2019,
Phys. Rev. D, {100}, 044019
[arXiv:1906.08742 [gr-qc]]

\bibitem[\protect\citeauthoryear{Cruz-Dombriz \& Saez-Gomez}{2012}]{reviews5}
de la Cruz-Dombriz A., Saez-Gomez D., 2012,
  Entrp, {14}, 1717
  [arXiv:1207.2663 [gr-qc]].
 
\bibitem[\protect\citeauthoryear{Dimopoulos}{2021}]{dimo} Dimopoulos K., 2021, \textit{Introduction to Cosmic Inflation and Dark Energy}, CRC Press

\bibitem[\protect\citeauthoryear{Doneva et al.}{2013}]{Doneva:2013qva}
Doneva D.D., Yazadjiev S.S., Stergioulas N., Kokkotas K.D., 2013,
Phys. Rev. D, {88}, 084060
[arXiv:1309.0605 [gr-qc]]

\bibitem[\protect\citeauthoryear{Horbatsch et al.}{2015}]{Horbatsch:2015bua}
Horbatsch M., Silva H.O., Gerosa D., Pani P., Berti E.,
Gualtieri L., Sperhake U., 2015,
Class. Quant. Grav., {32}, 204001
[arXiv:1505.07462 [gr-qc]]

\bibitem[\protect\citeauthoryear{Katsuragawa et al.}{2022}]{Katsuragawa:2022} Numajiri K., Katsuragawa T., Nojiri S., 2022, PLB, 826, 136929

\bibitem[\protect\citeauthoryear{Kilic, et al.}{2021}]{Cilic} Kilic M., Bergeron P., Blouin S., Bedard A., 2021, MNRAS, 503, 5397

\bibitem[\protect\citeauthoryear{Lobato et al.}{2020}]{Lobato:2020fxt}
Lobato R., Louren\c{c}o O., Moraes P.H.R.S., Lenzi C.H., de
Avellar M., de Paula W., Dutra M., Malheiro M., 2020,
JCAP, {12}, 039 
[arXiv:2009.04696 [astro-ph.HE]]

\bibitem[\protect\citeauthoryear{McDonald, et al.}{2006}]{McDonald} McDonald P. et al., 2006, ApJS, {163}, 80

\bibitem[\protect\citeauthoryear{Motahar et al.}{2017}]{Motahar:2017blm}
Motahar Z., Bl\'azquez-Salcedo J.L., Kleihaus B.,
Kunz J., 2017,
Phys. Rev. D, {96}, 064046
[arXiv:1707.05280 [gr-qc]]

\bibitem[\protect\citeauthoryear{Naf \& Jetzer}{2010}]{Naf} Naf J., Jetzer P., 2010, Phys. Rev. D, 81, 104003 [arXiv:1004.2014 [gr-qc]]
 
\bibitem[\protect\citeauthoryear{Nojiri, Odintsov \& Oikonomou}{2017}]{reviews4}
 Nojiri S., Odintsov S.D., Oikonomou V.K., 2017,
  Phys.\ Rept., {692}, 1
  [arXiv:1705.11098 [gr-qc]]

\bibitem[\protect\citeauthoryear{Nojiri \& Odintsov}{2011}]{book}
Nojiri S., Odintsov S.D., 2011,
   Phys.\ Rept.,  {505}, 59
   [arXiv:1011.0544 [gr-qc]]

\bibitem[\protect\citeauthoryear{Nojiri \& Odintsov}{2003}]{Nojiri:2003ft}
Nojiri S., Odintsov S.D., 2003,
PhRvD, {68}, 123512 [arXiv:hep-th/0307288]

\bibitem[\protect\citeauthoryear{Oikonomou}{2021}]{Oikonomou:2021iid}
Oikonomou V.K., 2021,
Class. Quant. Grav., {38}, 175005
[arXiv:2107.12430 [gr-qc]]

\bibitem[\protect\citeauthoryear{Odintsov \& Oikonomou}{2021}]{Odintsov:2021qbq}
Odintsov S.D., Oikonomou V.K., 2021,
Phys. Dark Univ., {32}, 100805
[arXiv:2103.07725 [gr-qc]]

\bibitem[\protect\citeauthoryear{Olmo}{2011}]{reviews6}
Olmo G.J., 2011,
  IJMPD, {20}, 413
  [arXiv:1101.3864 [gr-qc]].
\bibitem[\protect\citeauthoryear{Olmo, Rubiera-Garcia \& Wojnar}{2020}]{Wojnar3} Olmo G.J., Rubiera-Garcia D., Wojnar A., 2020, Phys. Rept., 876, 1 [arXiv:1912.05202 [gr-qc]]

\bibitem[\protect\citeauthoryear{Pani \& Berti}{2014}]{Pani:2014jra}
Pani P., Berti E., 2014,
Phys. Rev. D, {90}, 024025
[arXiv:1405.4547 [gr-qc]]

\bibitem[\protect\citeauthoryear{Panotopoulos, et al.}{2021}]{Panotopoulos:2021sbf}
Panotopoulos G.,, Tangphati T., Banerjee A., Jasim M.K.,
[arXiv:2104.00590 [gr-qc]]

\bibitem[\protect\citeauthoryear{Perlmutter et al.}{1999}]{Perlmutter} Perlmutter S. {et al.}  [Supernova Cosmology Project Collaboration], 1999, ApJ, {517}, 565 [arXiv:astro-ph/9812133]

\bibitem[\protect\citeauthoryear{Riess et al.}{1998}]{Riess1} Riess A.G. {et al.}  [Supernova Search Team Collaboration], 1998, AJ, {116}, 1009 [arXiv:astro-ph/9805201]

\bibitem[\protect\citeauthoryear{Riess et al.}{2004}]{Riess2} Riess A.G. {et al.}  [Supernova Search Team Collaboration], 2004, ApJ,  {607}, 665 [arXiv:astro-ph/0402512]

\bibitem[\protect\citeauthoryear{Sarmah, Kalita \& Wojnar}{2022}]{Wojnar1} Sarmah L., Kalita S., Wojnar A., 2022, PhRvD, 105, 024028 [arXiv:2111.08029 [gr-qc]]

\bibitem[\protect\citeauthoryear{Schimdt, et al.}{2007}]{Schmidt} C. Schimdt C. et al., 2007, A\&A, {463}, 405

\bibitem[\protect\citeauthoryear{Silva et al.}{2015}]{Silva:2014fca}
Silva H.O., Macedo C.F.B., Berti E., Crispino L.C.B., 2015,
Class. Quant. Grav., {32}, 145008
[arXiv:1411.6286 [gr-qc]]

\bibitem[\protect\citeauthoryear{Spergel et al.}{2003}]{Spergel} Spergel D.N. {et al.}  [WMAP Collaboration], 2003, ApJS,  {148}, 175 [arXiv:astro-ph/0302209]

\bibitem[\protect\citeauthoryear{Weinberg}{1989}]{Weinberg} Weinberg S., 1989, Rev. Mod. Phys., {61}, 1

\bibitem[\protect\citeauthoryear{Wojnar}{2021}]{Wojnar2}   Wojnar A., 2021, Int. J. Geom. Meth. Mod. Phys., 18, 2140006 [arXiv:2012.13927 [gr-qc]]























\end{thebibliography}
\end{document}